# Visualizing molecular unidirectional rotation

**Kang Lin[1], Qiying Song[1], Xiaochun Gong[1], Qinying Ji[1], Haifeng Pan[1], Jingxin Ding[1], Heping Zeng[1,2,†], and Jian Wu[1,‡]**

[1]*State Key Laboratory of Precision Spectroscopy, East China Normal University, Shanghai 200062, China*
[2]*Synergetic Innovation Center of Quantum Information and Quantum Physics, University of Science and Technology of China, Hefei, Anhui 230026, China*

[†] hpzeng@phy.ecnu.edu.cn
[‡] jwu@phy.ecnu.edu.cn

**Abstract:** A molecule can be optically accelerated to rotate unidirectionally at a frequency of a few terahertzes which is many orders higher than the classical mechanical rotor. Such a photon-induced ultrafast molecular unidirectional rotation has been well explored as a controllable spin of the molecular nuclear wave packet. Although it has been observed for more than 10 years, a complete imaging of the unidirectional rotating nuclear wave packet is still missing, which is essentially the cornerstone of all the exploring applications. Here, for the first time, we experimentally visualize the time-dependent evolution of the double-pulse excited molecular unidirectional rotation by Coulomb explosion imaging the rotational nuclear wave packet. Our results reveal comprehensive details undiscovered in pioneering measurements, which exhibits as a joint of the quantum revival of the impulsively aligned rotational wave packet and its unidirectional rotation following the angular momentum conservation. The numerical simulations well reproduce the experimental observations and intuitively revivify the thoroughgoing evolution of the rotational wave packet.

**Introduction**

Control the motion of a molecule by coherently manipulating the nuclear wave packet using ultrashort laser pulses is significant for a large variety of processes, ranging from extreme ultraviolet radiation generation[1-3], molecular orbital and structure imaging[4,5], to chemical reaction steering[6]. In addition to fixing the molecular axis in space[7], a molecule can be optically kicked to spin unidirectionally at terahertz frequency in the polarization plane of the excitation laser pulses[8-10]. Similar to the well-studied spin of an electron, the molecular unidirectional rotation (UDR) is the directional spin of the molecular nuclear wave packet which is coherently controllable using ultrashort laser pulses.

A molecule can be step-by-step accelerated to a very high rotation frequency of more than ten terahertzes by the optical centrifuge[11-14] of an energetic picosecond pulse of rotating linearly polarized laser field; or impulsively kicked to rotate at a frequency of few terahertzes by the double-pulse[15-18] or chiral-multipulse[19,20] excitation schemes. The extremely fast UDR of a molecule is observed to break molecular bond[11,12], rotationally Doppler or Raman shift the frequency of a circularly polarized light dependent on the rotation sense[14,17], boost the ionization of particular electronic state[15,19], and resist collisions[21]. The molecular UDR is also interesting to defect molecules[22], generate molecular vortices[23], and work as molecular waveplate[24]. However, a comprehensive picture of the time-dependent evolution of the molecular UDR, which essentially serves as the basis of all the initiated applications, has lacked for more than 10 years since its observation. On the one hand, the molecular UDR has been mostly observed in the frequency domain using a spectral narrowband laser pulse[14] to well resolve the frequency dependence and consequently limited temporal resolution. On the other hand, the rotational Doppler shift of the carrier frequency of an ultrashort laser pulse[17] provides an adequate temporal resolution. Nevertheless, it is hard to trace the full evolution of the molecular UDR since its observability demands a strong molecular anisotropy.

Here we overcome these hurdles by direct Coulomb explosion imaging[25,26] the rotational nuclear wave packet of a molecule following its strong field ionization, providing a temporal resolution in femtosecond range with no requisite molecular anisotropy. Our experiments as well as the numerical simulations comprehensively visualize the time-dependent evolution of the rotational wave packet of a unidirectionally rotating molecule. It, for the first time, intuitively presents us a full picture of the molecular UDR, which strengthens our understanding of this fascinating quantum dynamics and motivates new applications.

**Results**

We create the molecular UDR by sequentially kicking the molecule with two ultrashort linearly polarized laser pulses[15-18], i.e. the alignment pulse at t=0 and rotation pulse at $t=t_r$, as schematically illustrated in Fig. 1. The instantaneous snapshots of the molecular nuclear wave packet at various time $t=t_p$ are taken by Coulomb exploding the molecule using a time-delayed circularly polarized (probe) pulse in a reaction microscope of COLTRIMS[27,28] ("COLd Target Recoil Ion Momentum Spectroscopy" see Methods). We characterize the evolution of the rotational wave packet by tracing its angular distribution in the polarization plane of the laser pulses, i.e. the azimuth angle $\phi$ between the molecular axis and z-axis in y-z plane.

Figure 2a displays the measured time-dependent angular distribution of $N_2$ retrieved from the double-ionization-induced Coulomb explosion channel of $N_2 + n\hbar\omega \rightarrow N^+ + N^+ + 2e$, denoted as ($N^+$, $N^+$). The arrivals of the alignment and rotation pulses are indicated by the bright lines at time of t=0 and 8.6 ps, respectively. To increase the visibility and eliminate the bias induced by the non-perfect circularity of the probe pulse, we firstly normalize the total probability of the angular distribution to unity for each time delay and then subtract the averaged angular distribution before the arriving of the first pulse.

Figure 2b illustrates the normalized angular distribution which presents the full evolution of the rotational nuclear wave packet excited by two time-delayed ultrashort linearly polarized pulses. Figure 2c shows the numerically simulated time-dependent angular distribution (see Methods), which very well agrees with the experimental measurement. The tilted angular distribution after the kick of the rotation pulse directly illustrates the UDR of the molecule. The animated dynamics evolutions of the measured angular distribution (Supplementary Movie s1) and the numerically simulated three-dimensional rotational wave packet (Supplementary Movie s2) intuitively revivify the molecular UDR. To the best of our knowledge, it is the first full picture of the time-dependent evolution of the rotational nuclear wave packet of an UDR molecule.

**Discussion**

The molecule firstly aligns to the polarization direction of the alignment pulse with an intense angular distribution around $\phi$=0º and ±180º at t=0.2 ps. The impulsively excited rotational wave packet field-free evolves to alignment and anti-alignment at the (fractional) revivals owing to the time-dependent phase beating of different rotational states[29-31]. Figure 3a depicts the rapid quantum transformation of the rotational wave packet from alignment to anti-alignment (~500 fs) around the half revival time of t=0.5 $t_{rev}$, where $t_{rev}$ = 8.4 ps is the full revival time of $N_2$. Figure 3b shows the polar plots of the angular distribution at t=$t_1$, $t_2$, and $t_3$ indicated in Fig. 3a. The corresponding three-dimensional rotational wave packets from our numerical simulations (see Methods) are presented in Figs. 3c-e, which agree well with the experiments and intuitively show the quantum evolution of the rotational wave packet. The rotational wave packet at alignment maximum is cigar-shaped (Figs. 3c & d) while it has a disk-shaped distribution (Fig. 3e) at anti-alignment maximum[32]. It evolves as a function of time but the angular distribution always mirrors with respect to the polarization vector of the alignment pulse.

A second ultrashort laser pulse polarized at 45º in the y-z plane is applied when the rotational wave packet is aligned along the z-axis at t = 8.6 ps. This impulsive torque unidirectionally rotates the cigar-shaped rotational wave packet to a frequency of terahertz[17,18]. Figure 4a depicts the detailed evolution of the angular distribution around t = 1.5 $t_{rev}$, and the corresponding polar plots at t=$t_1$, $t_2$, and $t_3$ are shown in Fig. 4b. As compared to that around t= 0.5 $t_{rev}$ (Fig. 3a), the angular distribution in Fig. 4a tilts as a function of time which directly illustrates the UDR of the molecule. The nuclear wave packet clockwise rotates and meanwhile evolves from alignment to anti-alignment. As visualized in Figs. 4c-e, this joint motion leads to a rotating-cigar or rotating-disk of the rotational nuclear wave packet, agreeing well with the experimentally imaged angular distribution of the molecule in *y-z* plane as shown in Fig. 4b. Due to the weak molecular anisotropy at the anti-alignment revival, this rotational wave packet of rotating-disk was not observed in recent experiments[17,18] based on the rotational Doppler shift of the carrier frequency of a circularly polarized pulse.

We notice that the molecular UDR is hard to be traced by watching the temporal evolution of $<\cos^2\phi>$, which is well-used to qualify the degree of molecular alignment[33]. The average of $<\cos^2\phi>$ = 0.5 indicates the random molecular orientation, while alignment and anti-alignment distributions result in values larger or smaller than 0.5, repectively[33,34]. As shown in Fig. 5a, no noticeable difference is observed for $<\cos^2\phi>$ before and after the kick of the rotation pulse at t = 8.6 ps. The $<\cos^2\phi>$ is little larger than 0.5 at t<0 owing to the bias of the non-perfect circularity of the probe pulse which is not eliminated here. It is predicted that the molecular spinning can confine the molecule in the plane spanned by the laser field vectors of the alignment and rotation pulses[9], which is indicated by the constant drop of the ($N^+$, $N^+$) yield along *x*-axis at t > 8.6 ps, i.e. after the kick of the rotation pulse, as illustrated in Fig. 5b. The decrease of the yield between 0 < t < 8.6 ps might be originated from permanent molecular alignment induced by the alignment pulse.

To demonstrate the robustness of our technique and test the validity of the data analysis, we adjusted the time delay of the 45º polarized rotation pulse to t = 8.08 ps when the molecule is anti-aligned with a disk-shaped nuclear wave packet. Figures s1 & s2 in the Supplementary Information depict the corresponding results. The numerical simulations agree well with the experimental measurements. As compared to Fig. 2, the tilting of the time-dependent angular distribution mirrors, indicating reversed rotation direction of the molecule. It agrees with the recent rotational Doppler frequency shift measurements[17,18], and much more details of the rotational wave packet evolution are observed here. The rotation pulse kicks the molecule anti-aligned in the plane perpendicular to the vector of the alignment pulse and counterclockwise rotates the disk-shaped nuclear wave packet, which afterwards can evolve into rotating cigar owing to the phase beating of the populated rotational states.

Alternatively, counterclockwise rotating nuclear wave packet can be created by kicking the aligned molecule at t = 8.6 ps using a rotation pulse polarized along - 45º with respect to *z*-axis in the *y-z* plane. The corresponding results are displayed in Figs. s3 & s4 in the Supplementary Information. Interestingly, both experimentally and numerically, an entire increase (Figs. 2b & c and Supplemental Figs. s1a & b) or decrease (Supplemental Figs. s2a & b) of the angular distribution is observed over all the time after the kick of the rotation pulse. Regardless of the rotation sense of the molecule, the shift of the angular distribution solely depends on the polarization direction of the rotation pulse. It might be due to the Stark effect induced by the rotation pulse which swings the rotational wave packet to its polarization direction.

In conclusion, for the first time, a comprehensive evolution of the double-pulse excited rotational wave packet of an UDR molecule is directly visualized using laser Coulomb explosion imaging. Our findings reveal a joint dynamics of the periodical quantum revivals of an impulsively aligned rotational wave packet and its unidirectional rotation governed by the angular momentum conservation. By kicking the molecule at the alignment or anti-alignment revivals of an impulsively aligned molecule, a cigar- or disk-shaped nuclear wave packet is spun to rotate clockwise or counterclockwise, which afterwards involves into rotating cigar and disk owing to the quantum beating of the populated rotational states of different energies. These full dynamics are very well intuitively reproduced by the numerical simulations by solving the time-dependent Schrödinger equation. The thoroughgoing knowledge of the evolution of the field-free molecular UDR is anticipated to bring a great leap forward in research on molecular optical physics.

## Methods

**Experimental technique.** Figure 1 schematically illustrates the experimental setup. A linearly polarized femtosecond laser pulse (25 fs, 790 nm, 10 kHz) produced from a multipass amplifier Ti:sapphire laser system is split into three parts, denoted as the alignment pulse, rotation pulse, and probe pulse, respectively. The polarization of the rotation pulse is rotated by 45º with respect to the alignment pulse, and the probe pulse is adjusted to circular polarization. The three pulses are afterwards recombined with adjustable time delays and focused onto a supersonic $N_2$ molecular beam in an ultrahigh vacuum chamber of COLTRIMS[27, 28] setup, where the molecule is spun to rotate unidirectionally and the rotational nuclear wave packet evolution is imaged using strong-field Coulomb explosion imaging.

The molecule is impulsively aligned by the alignment pulse, which is then kicked to rotate unidirectionally by the rotation pulse. An intense circularly polarized probe pulse multiply ionizes and consequently Coulomb explodes the molecule to image the instantaneous distribution of the nuclear wave packet in the polarization plane. The peak intensities of the alignment, rotation, and probe pulses in the interaction region are measured to be $5\times10^{13}$, $5\times10^{13}$, and $1.3\times10^{15}$ W/cm$^2$, respectively. The temporal duration of the alignment and rotation pulses are stretched to ~50 fs, while the probe pulse is ~25 fs. The ionization created fragment ions are guided by a weak homogeneous static electric field (~10 V/cm) to be detected by a time- and position-sensitive microchannel plate detector at the end of the spectrometer. The three-dimensional momenta of the fragment ions are retrieved from the measured time-of-flight and positions. We identify the fragment ions ejected from the same molecule based on the two-particle correlation. The extremely fast Coulomb explosion allows us to directly reconstruct the molecular axis from the relative momentum of the correlated fragment ions and hence the distribution of the rotational nuclear wave packet.

Since the $N_2$ molecule is mostly ionized and fragmentizes when the molecular axis is parallel to the laser field vector, we restrict our data analysis for molecules in the polarization plane of the circularly polarized probe pulse in which the molecule is impulsively kicked to rotate unidirectionally. By step-by-step scanning the time delay of the probe pulse, a serial of snapshots are taken to retrieve the full time-dependent evolution of the rotational wave packet. We note that the Coulomb explosion imaging is insensitive to the molecular anisotropy and the time resolution is comparable to the temporal duration of the probe pulse of tens femtoseconds in our experiments.

**Numerical simulation.** We model the UDR of a molecule by numerically solving the time-dependent Schrödinger equation[29,35] $i\hbar\partial|\psi\rangle/\partial t = H_{eff}|\psi\rangle$ for the rotational state $|\psi\rangle = \Sigma_{J,M} C_{JM}|J,M\rangle$, where $H_{eff} = B_0 J(J+1) - 0.5\Delta\alpha\sin^2\theta(\varepsilon_z^2\cos^2\phi + \varepsilon_y^2\sin^2\phi + 2\varepsilon_z\varepsilon_y\cos\phi\sin\phi)$ is the effective Hamiltonian, $B_0$ is the molecular rotational constant, $\Delta\alpha$ is the polarizability difference between the components parallel and perpendicular to the molecular axis, $\theta$ and $\phi$ are the polar and azimuth angles of the molecular axis with respect to the $x$- and $z$-axes, and $\varepsilon_y$ and $\varepsilon_z$ are the envelopes of the laser field vector along the $x$- and $z$-axes, respectively. We first calculated the term $P_{J_0M_0}(\theta, \phi, t) = |\Sigma_{J,M} C_{JM}(t)Y_{JM}(\theta, \phi)|^2$ for each initial molecular rotational state $|\psi(t=0)\rangle_{J_0M_0} = |J_0, M_0\rangle$, where $Y_{JM}(\theta, \phi)$ is the spherical harmonic functions. We then assembled them by considering the temperature-dependent Boltzmann distribution of the initial rotational states, and eventually obtained the time-dependent probability density distribution of the rotational wave packet $P(\theta, \phi, t)$.

In the numerical simulation, the molecular parameters of $N_2$ are $B_0$=1.98 cm$^{-1}$, $\Delta\alpha$ = 1.0 Å$^3$. The initial rotational temperature of the $N_2$ molecule is set to be 20 K, which is determined by matching the calculated arriving time of the first alignment maximum as well as other revivals after the excitation of the alignment pulse to the experiments. The parameters of the alignment and rotation pulses are chosen similar to the experiments. By coupling the rotational states of $|J, M\rangle$ and $|J\pm 2, M\rangle$, the molecule is aligned by the alignment pulse, while the rotation pulse further couples different M states and spins the rotational wave packet. As shown in Supplemental Fig. s5 in the Supplementary Information, the angular momentum of the rotational wave packet along $x$-axis $\langle J_x\rangle$ gradually increases from zero to 0.39 or -0.33 during the kick of the rotation pulse, which is constant after the end of the laser pulse owing to the angular momentum conservation. The larger amplitude of $\langle J_x\rangle$ when the rotation pulse matches the alignment maximum than the anti-alignment maximum is consistent with the fact that the cigar-shaped rotational wave packet is more localized along the $z$-axis as compared to the disk-shaped one in the $y$-$z$ plane.

*Phys. B: At. Mol. Opt. Phys.* **41,** 074018 (2008).

33. Litvinyuk, I. V., Lee, K. F., Dooley, P. W., Rayner, D. M., Villeneuve, D. M. & Corkum, P. B. Alignment-dependent strong field ionization of molecules. *Phys. Rev. Lett.* **90,** 233003 (2003).

34. Goban, A., Minemoto, S. & Sakai, H. Laser-field-free molecular orientation. *Phys. Rev. Lett.* **101,** 013001 (2008).

35. Wu, J. & Zeng, H. Field-free molecular orientation control by two ultrashort dual-color laser pulses, *Phys. Rev. A* **81**, 053401 (2010).

<>
**Acknowledgments** We acknowledge helpful discussions with Ilya Sh. Averbukh, O. Faucher, and C. Hang. This work is supported by the "Eastern Scholar" Program, the NCET in University (NCET-12-0177), project from the SSTC (13QH1401400), the "ShuGuang" project (12SG25), and the National Natural Science Fund (11374103).

**Author Contributions** All authors contributed to the experiment and analysis. J.W. and K. L. contributed the theoretical simulation. All authors discussed the results and commented on the manuscript.

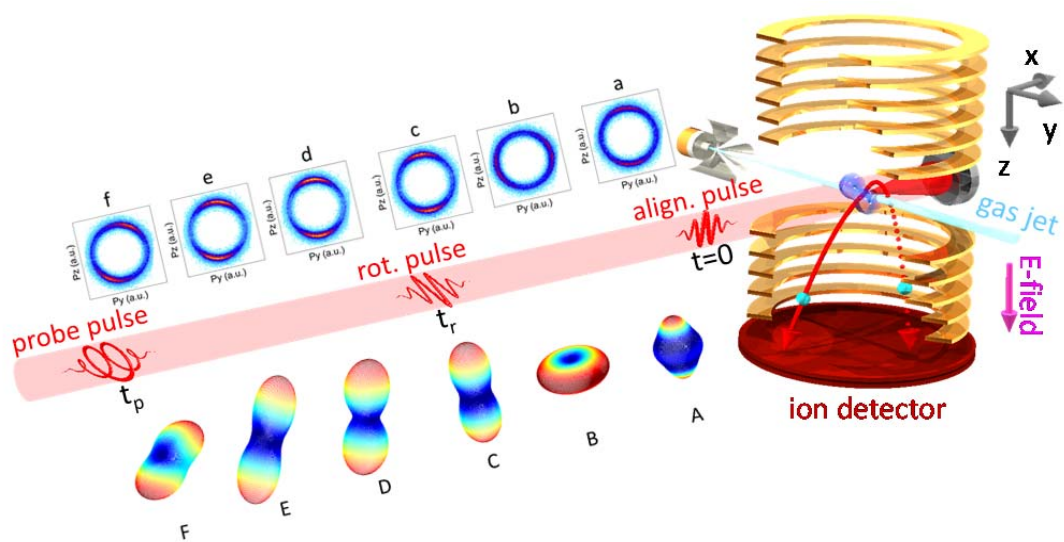

**Figure 1 | Experimental setup.** The measurement is performed in an ultrahigh vacuum apparatus of COLTRIMS. A nitrogen molecule contained in a supersonic gas jet is impulsively aligned by an alignment pulse, which is subsequently kicked to rotate unidirectionally by a properly matched rotation pulse. An intense circularly polarized probe pulse with adjustable time delay is used to Coulomb explosion image the evolution of the rotational wave packet. The insets a-f above the laser beam are the snapshots of the measured momentum distribution of the fragments of the Coulomb exploded nitrogen molecule; while the insets A-F below the laser beam are the simulated distributions of the rotational wave packet at various time delays.

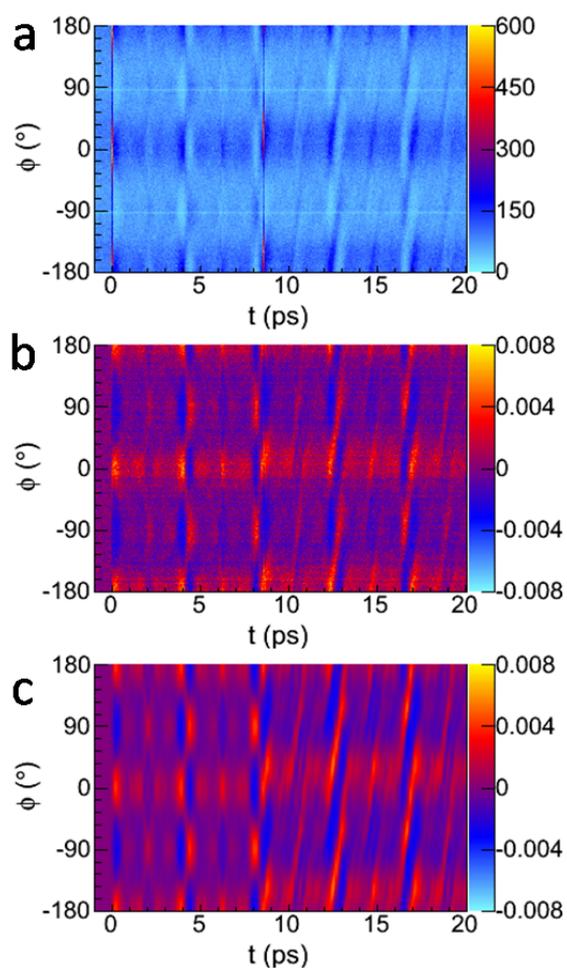

**Figure 2 | Time-dependent angular distribution.** (**a**) Count plot of experimentally measured angular distribution of ($N^+$, $N^+$) fragments in *y-z* plane ejected from a doubly ionized nitrogen molecule as a function of the time delay of the probe pulse. (**b**) The normalized angular distribution of (**a**). (**c**) Numerically simulated angular distribution of the rotational wave packet in *y-z* plane spanned by the field vectors of the alignment and rotation pulses. The 45º polarized rotation pulse arrives at the maximum of alignment revival at t=8.6 ps.

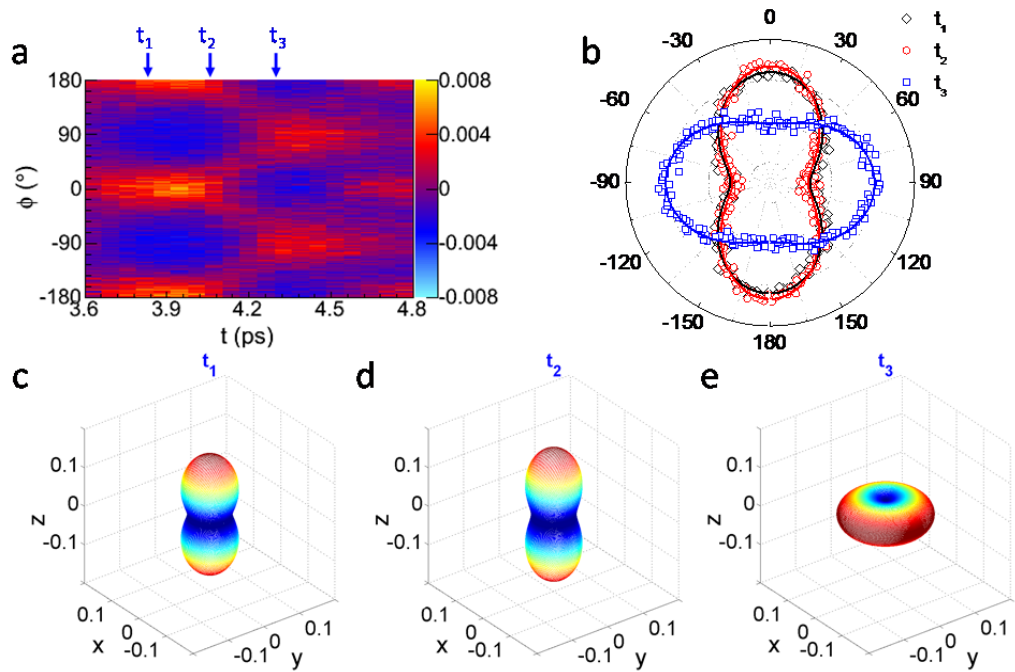

**Figure 3 | Angular distribution and rotational wave packet around t=0.5 $t_{rev}$.** (**a**) Normalized angular distribution around t=0.5 $t_{rev}$ taken from Fig. 2(**b**). (**b**) Polar plots of the angular distribution at t=$t_1$, $t_2$, and $t_3$ as indicated in (**a**). (**c-e**) Numerically simulated rotational wave packet distributions at t=$t_1$, $t_2$, and $t_3$. The rotational wave packet evolves from cigar-shaped alignment to disk-shaped anti-alignment due to the sole excitation of the alignment pulse.

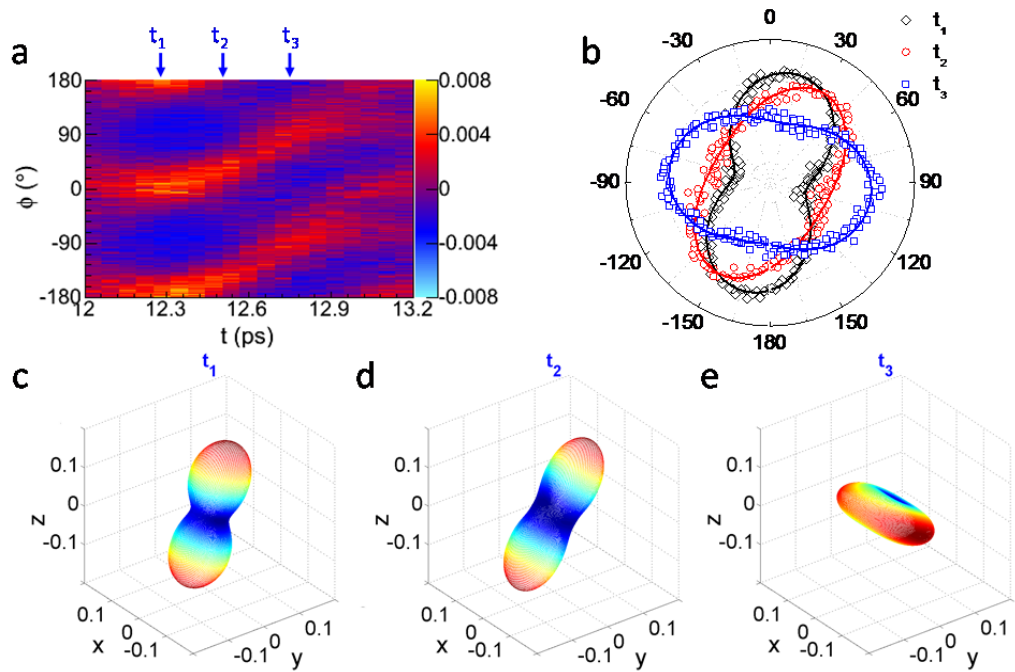

**Figure 4 | Angular distribution and rotational wave packet around $t=1.5\, t_{rev}$. (a)** Normalized angular distribution around $t=1.5\, t_{rev}$ taken from Fig. 2(**b**). (**b**) Polar plots of the angular distribution at $t=t_1$, $t_2$, and $t_3$ as indicated in (**a**). (**c-e**) Numerically simulated rotational wave packet distributions at $t=t_1$, $t_2$, and $t_3$. The rotational wave packet transforms from cigar to disk and meanwhile clockwise rotates due to the sequential kicks of the alignment and rotation pulses.

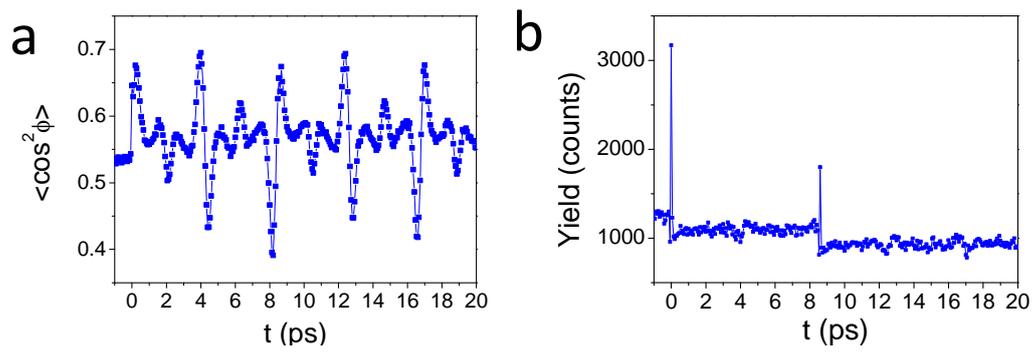

**Figure 5 | Time-dependent alignment degree and yield.** (**a**) Measured $\langle\cos^2\phi\rangle$ in the *y-z* plane, and (**b**) yield of the $(N^+, N^+)$ fragments along the *x*-axis as a function of the time delay of the probe pulse.

*Supplementary Information*

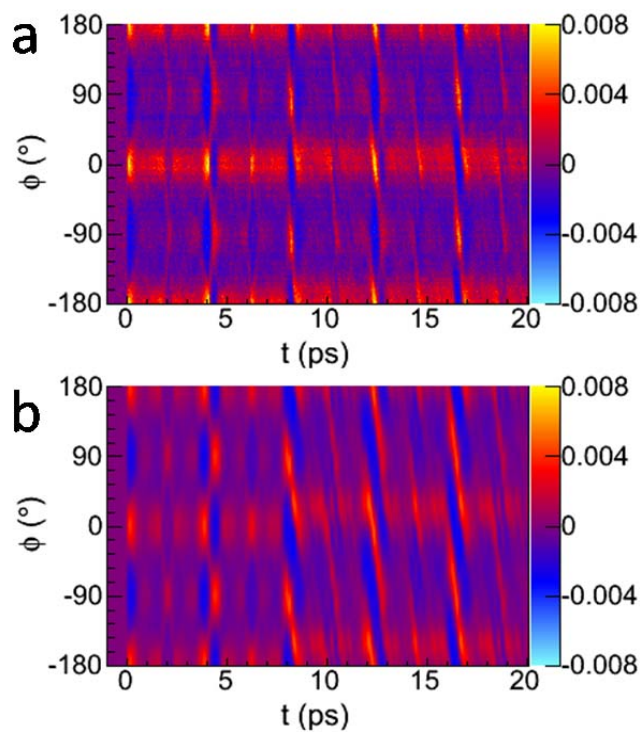

**Supplementary Figure s1 | Time-dependent angular distribution.** (**a**) Measured density plot of normalized angular distribution of ($N^+$, $N^+$) fragments in *y-z* plane as a function of the time delay of the probe pulse. (**b**) Numerically simulated angular distribution of the rotational wave packet in *y-z* plane spanned by the field vectors of the alignment and rotation pulses. The 45º polarized rotation pulse arrives at the maximum of anti-alignment revival at t=8.08 ps. The other conditions are the same as Fig. 2. The time-dependent angular distribution oppositely tilts as compared to Fig. 2, indicating reversed rotation direction of the nitrogen molecule.

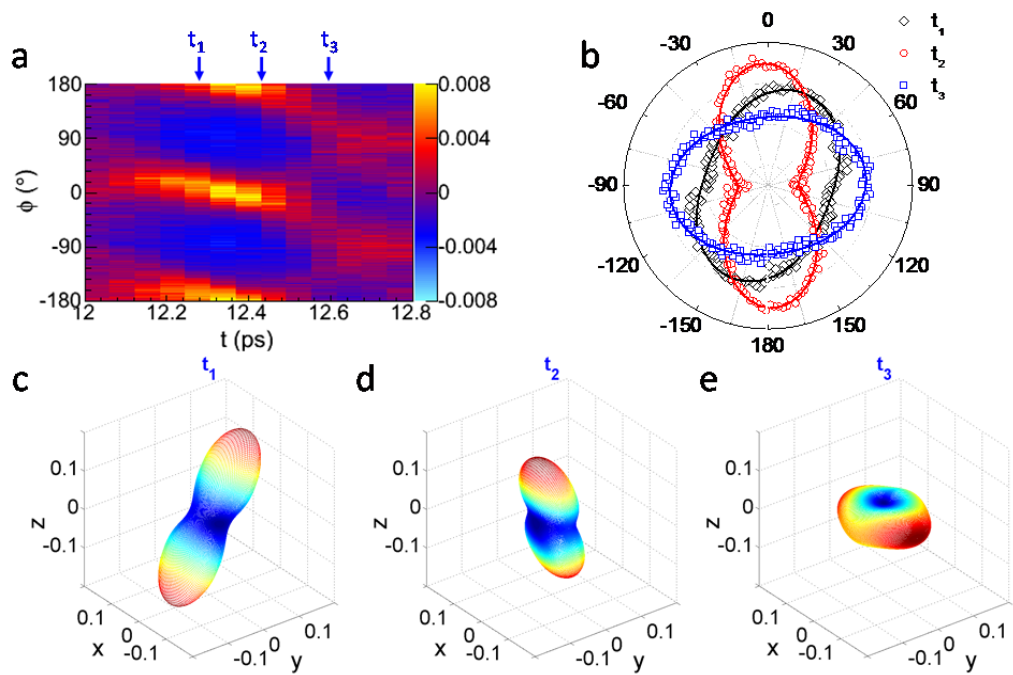

**Supplementary Figure s2 | Angular distribution and rotational wave packet around t=1.5 $t_{rev}$.** (**a**) Normalized angular distribution around t=1.5 $t_{rev}$ taken from Fig. s1(**a**). (**b**) Polar plots of the angular distribution at t=$t_1$, $t_2$ and $t_3$ as indicated in (**a**). (**c-e**) Numerically simulated rotational wave packet distributions at t=$t_1$, $t_2$ and $t_3$. The rotational wave packet transforms from cigar to disk and meanwhile counterclockwise rotates due to the sequential kicks of the alignment and rotation pulses.

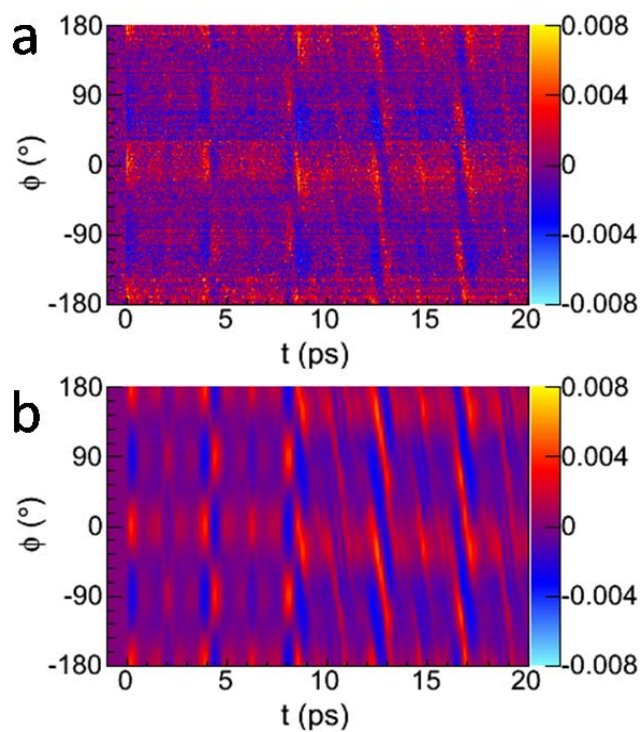

**Supplementary Figure s3 | Time-dependent angular distribution.** (**a**) Measured density plot of normalized angular distribution of $(N^+, N^+)$ fragments in *y-z* plane as a function of the time delay of the probe pulse. (**b**) Numerically simulated angular distribution of the rotational wave packet in *y-z* plane spanned by the field vectors of the alignment and rotation pulses. The -45° polarized rotation pulse arrives at the maximum of alignment revival at t=8.6 ps. The time-dependent angular distribution oppositely tilts as compared to Fig. 2, indicating reversed rotation direction of the nitrogen molecule.

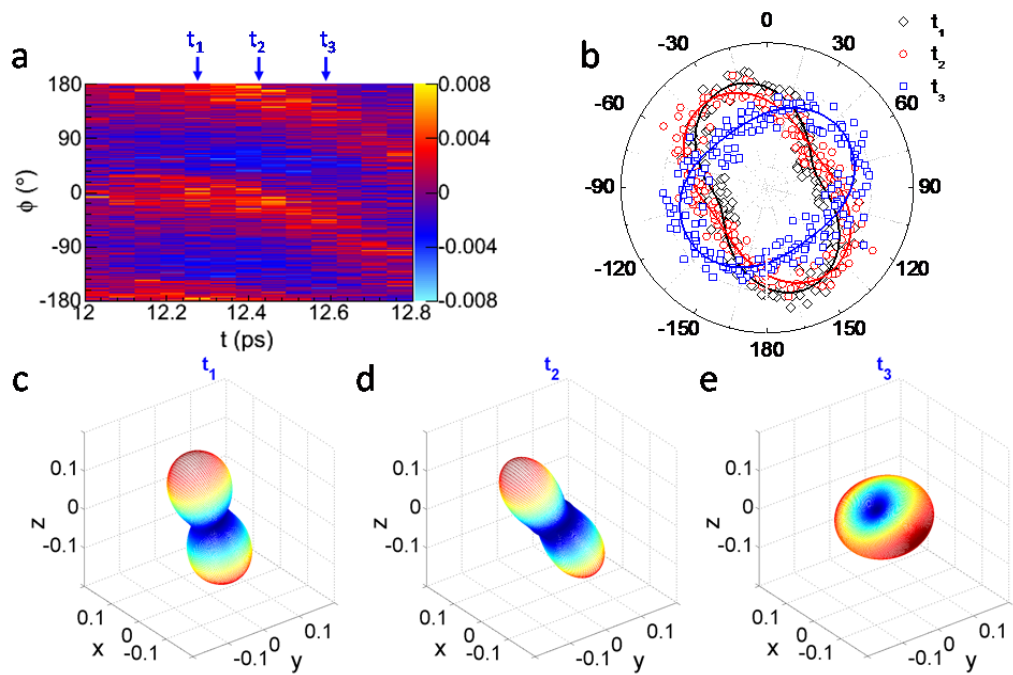

**Supplementary Figure s4 | Angular distribution and rotational wave packet around t=1.5 $t_{rev}$.** (**a**) Normalized angular distribution around t=1.5 $t_{rev}$ taken from Fig. s3(**a**). (**b**) Polar plots of the angular distribution at t=$t_1$, $t_2$ and $t_3$ as indicated in (**a**). (**c-e**) Numerically simulated rotational wave packet distributions at t=$t_1$, $t_2$ and $t_3$. The rotational wave packet transforms from cigar to disk and meanwhile counterclockwise rotates due to the sequential kicks of the alignment and rotation pulses.

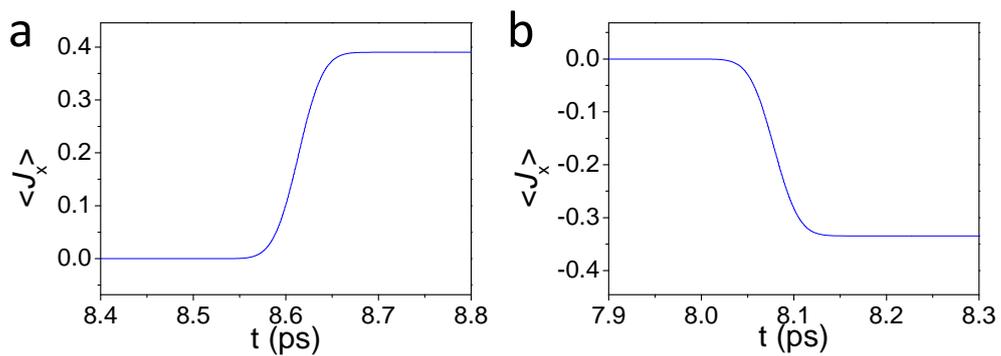

**Supplementary Figure s5 | Time-dependent angular momentum.** Numerically simulated time-dependent angular momentum $<J_x>$ around the arriving time of the rotation pulse. The rotation pulse gradually spins up the molecule to (**a**) clockwise ($<J_x>\ > 0$) and (**b**) counterclockwise ($<J_x>\ < 0$) rotate in the *y-z* plane when the rotation pulse arrives at the maxima of the (**a**) alignment and (**b**) anti-alignment revivals, respectively.

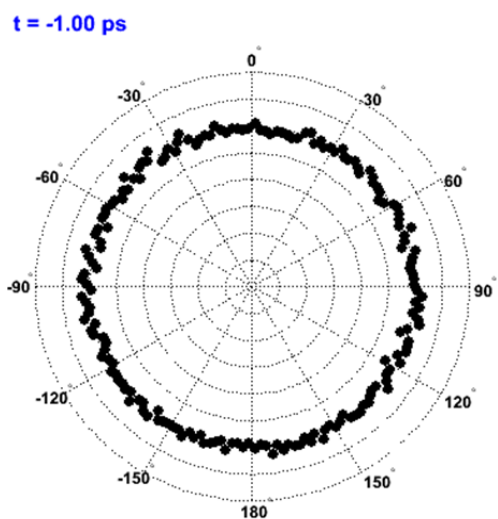

**Supplementary Movie s1 | Animated angular distribution.** Experimentally measured time-dependent evolution of the angular distribution of (N⁺, N⁺) fragments in *y-z* plane ejected from a doubly ionized nitrogen molecule.

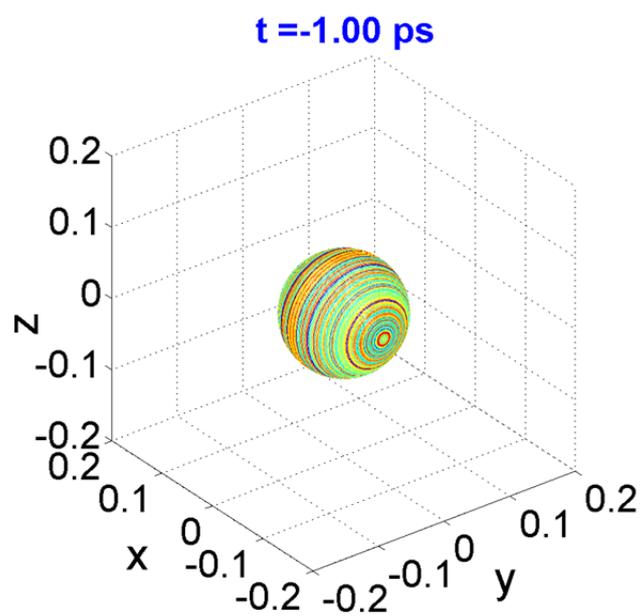

**Supplementary Movie s2 | Animated rotational wave packet.** Numerically simulated time-dependent evolution of three-dimensional rotational nuclear wave packet.